\documentclass[twocolumn, amssymb, amsmath, aps]{revtex4}
\usepackage{graphics}
\begin{document}

\title{Off-lattice Noise Reduced Diffusion-limited Aggregation in Three
Dimensions}

\author{Neill E. Bowler}
\email{Neill.Bowler@metoffice.com}
\affiliation{Met Office, Fitzroy Road, Exeter, EX1 3PB, UK}
\altaffiliation{Department of Physics, University of Warwick, Coventry, CV4 7AL, UK}

\author{Robin C. Ball}
\email{R.C.Ball@warwick.ac.uk}
\affiliation{Department of Physics, University of Warwick, Coventry, CV4 7AL, UK}

\begin{abstract}
Using off-lattice noise reduction it is possible to estimate the asymptotic
properties of diffusion-limited aggregation clusters grown in three
dimensions with greater accuracy than would otherwise be possible.
The fractal dimension of these aggregates is found to be $2.50\pm0.01$,
in agreement with earlier studies, and the asymptotic value of the
relative penetration depth is $\frac{\xi}{R_{dep}}=0.122\pm0.002$.
The multipole powers of the growth measure also exhibit universal asymptotes.
The fixed point noise reduction is estimated to be $\epsilon^{f}\sim0.0035$
meaning that large clusters can be identified with a low noise regime.
The slowest correction to scaling exponents are measured for a number
of properties of the clusters, and the exponent for the relative penetration
depth and quadrupole moment are found to be significantly different
from each other. The relative penetration depth exhibits the slowest
correction to scaling of all quantities, which is consistent with
a theoretical result derived in two dimensions. 
\end{abstract}
\maketitle

Diffusion-limited aggregation (DLA) \cite{witten} is an extensively
studied model of diffusion limited growth which appears to capture
the essential features of many different physical growth phenomena
\cite{langer,fujikawa,brady,nittmann}.
However, the fractals generated have evaded a complete understanding
for many years and there has recently been controversy over whether
diffusion-limited aggregates are truly fractal \cite{mandelbrot4}. 

DLA is modelled \cite{witten} by allowing particles to randomly walk
from a sphere surrounding the cluster, one at a time, until they contact
the cluster, at which point they are irreversibly stuck. Detailed
study \cite{somfai} has shown that DLA growth in two dimensions does
approach true fractal scaling, but with slowly decaying corrections
to scaling of the form \begin{equation}
Q_{N}=Q_{\infty}+CN^{-\nu}\label{correction to scaling}\end{equation}
 where $Q_{N}$ is some property of the cluster, tending towards the
value $Q_{\infty}$ as the number of particles in the cluster, $N$,
tends to infinity. Here the correction to scaling exponent $\nu$
is expected to exhibit some universality whilst the constant $C$
will not. For aggregates grown in two dimensions, it has been argued
\cite{somfai} that there should be no quantity whose correction to
scaling is slower than that of the relative penetration depth, $\frac{\xi}{R_{dep}}$,
where $R_{dep}$ is the average radius at which new particles are
deposited and $\xi$ is the standard deviation of the same. Here and
below we take as origin the centroid of the depositing particles.

When studying DLA grown on a lattice, reducing the shot noise associated
with the growth being by discrete particles \cite{tang} has proved
valuable in understanding the asymptotic properties of clusters, complicated
by their sensitivity to lattice anisotropy \cite{ball2,barker}.
Using a conformal map from the unit circle to the boundary
of a growing cluster, Hastings \& Levitov \cite{hastings} introduced
a technique for implementing a noise reduction scheme for DLA clusters
grown off-lattice. Rather than adding particles to the cluster, bumps
were added to the conformal map. Ball \emph{et al.\ {}}\cite{ball2002}
built on this work, allowing crescent-shaped bumps to be added to
DLA clusters without the need for a conformal map. In this approach,
the particles are allowed to diffuse normally until they contact the
cluster. At this point the new particle is touching the cluster, and
the distance between the centre of the new particle and the centre
of the particle it contacted in the cluster is equal to the diameter
of one particle. To implement noise reduction, this distance is reduced
by a factor $A<1$, so that the new particle is deposited partially
inside the cluster: the effect is to protrude a shallow bump of height
$A$ on the cluster perimeter. Since this method does not rely on
a conformal map it allows the growth of noise reduced DLA clusters
off-lattice in any dimension. 

Most of the work on DLA has been restricted to two dimensions. Meakin
pioneered work on DLA in higher dimensions, growing clusters in dimensions
up to $d=8$ \cite{tolman}. Much of that work has focused on estimating
the fractal dimension of DLA clusters and the scaling of the relative
penetration depth \cite{meakin2,meakin5,sander3}, yet firm conclusions
have proved difficult. There has also been some progress measuring
the multifractal spectrum of DLA in three dimensions \cite{schwarzer1,schwarzer2,gagne}.
However, Davidovitch \emph{et al.\ {}}\cite{jensen} recently claimed
that all previous attempts to measure $f(\alpha)$ in two dimensions
are poorly converged, so the early three dimensional measurements
should be taken with caution. Other work has also examined DLA on
a cubic lattice in the limit of zero noise \cite{batchelor}, and
the extension of the fixed scale transformation to 3 dimensions \cite{vespignani4}. 

\begin{figure}
\resizebox*{\columnwidth}{0.8 \columnwidth}{
\includegraphics{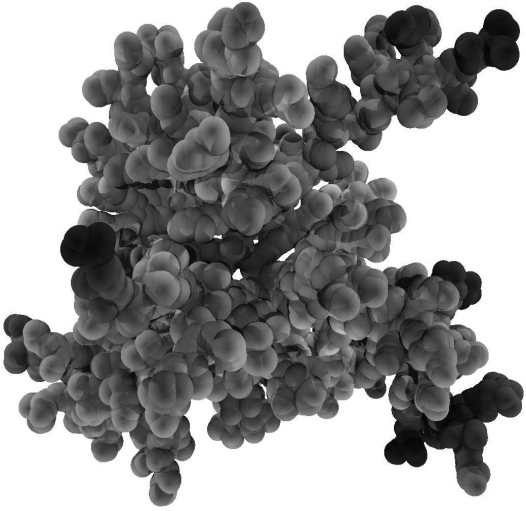}}
\caption[A DLA cluster grown off-lattice in three dimensions with $N=10^{4}$
particles and noise reduction factor $A=0.1$.]{A DLA cluster
grown in three dimensions with $N=10^{4}$ particles and noise reduction
factor $A=0.1$, where the different shading indicates a different
time of deposition on the cluster.}
\label{cluster in 3d}
\end{figure}

In this paper we exploit the new noise reduction techniques to explore
the convergence to scaling of DLA in three dimensions. We grew 1000
DLA clusters off-lattice in three dimensions spanning five different
values of the noise reduction $A=1$, $0.3$, $0.1$, $0.03$, $0.01$,
and an example cluster with $N=10^{4}$ particles and $A=0.1$ is
shown in figure \ref{cluster in 3d}. At regular intervals, the growth
of the clusters was suspended and $10^{5}$ probe particles were {}``fired
at'' the cluster: these probe particles were allowed to diffuse freely,
one at a time, until they contacted the cluster, at which point their
location was recorded and the particle was deleted. In this way the
growing properties of the clusters (such as the penetration depth
and multipole moments) were estimated. The code used is a direct descendent
of that of Meakin \cite{meakin2}, which in turn builds on the computational
tricks of Ball \& Brady \cite{ball} to speed up computation. Whilst
the code is truly lattice free, the smallest step size that a particle
was allowed to take was set equal to one particle radius. Comparisons
in 2 dimensions between this and an algorithm which uses a much smaller
minimum step size have shown that any effect that this has on the
results is the same order as the noise in the measurements attributable
to inter-cluster variability \cite{bowler2001}.

\section{Fractal dimension and correction to scaling exponents}

\begin{figure}
\begin{center}
\resizebox*{\columnwidth}{0.717 \columnwidth}{
\includegraphics{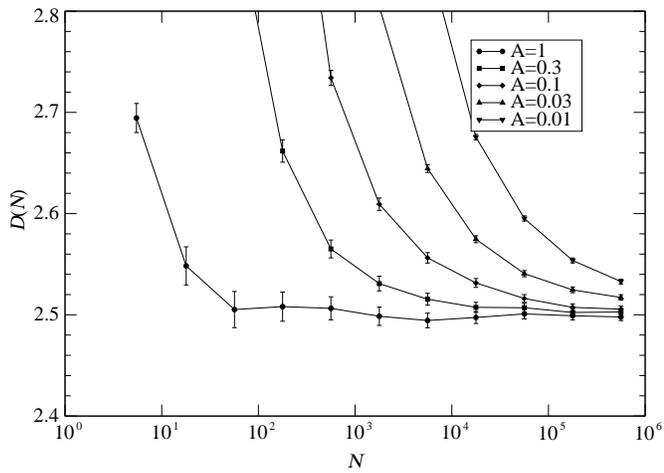}}
\end{center}
\caption[The measured fractal dimension of clusters grown in 3D.]{The measured
fractal dimension of DLA clusters, estimated by taking the local slope
of $R_{dep}$. The dimension appears to be converging to a dimension
of $D=2.50\pm0.01$, universal with respect to value of noise reduction
$A$.}
\label{fractal dimension 3d}
\end{figure}

We calculate the effective value of fractal dimension from the local
slope of the average radius of deposition vs number of particles $N$
according to \begin{equation}
D=\frac{\ln(N_{2})-\ln(N_{1})}{\ln(R_{dep}(N_{2}))-\ln(R_{dep}(N_{1}))}\label{local dimension}\end{equation}
 where the properties are measured at two different cluster sizes
$N_{1}$ and $N_{2}$. For the clusters grown, the value of the fractal
dimension is shown in figure \ref{fractal dimension 3d}. The dimensions
estimated for each value of $A$ appear to be converging to a common
value of $D=2.50\pm0.01$ which is consistent with previous computational
estimates \cite{tolman}. The data for $A=0.03$ and $A=0.01$ are
less well converged, and the results could be made more accurate with
data for larger $N$. 

We now consider the correction to scaling exponents for DLA in three
dimensions. In two dimensions it has been shown that no property of
the cluster has a slower correction to scaling than the relative penetration
depth and suggested that all properties should show influence of this
slowest correction \cite{somfai}. To measure the slowest correction
to scaling exponent of a quantity, we used differential plots which
proved effective for DLA in two dimensions \cite{ball2002}. For some
converging quantity $Q_{N}$, which displays a single correction to
scaling (eq.\ \ref{correction to scaling}), then
\begin{equation}
\frac{dQ_{N}}{d\ln(N)}=-\nu(Q_{N}-Q_{\infty}).
\end{equation}
 so a plot of $\frac{dQ}{d\ln(N)}$ against $Q$ should exhibit a
straight line with slope $-\nu$, intercepting the x-axis at the asymptotic
value $Q_{\infty}$. The differential is approximated by
\begin{equation}
\frac{dQ}{d\ln(N)} \simeq \frac{Q_{N_{2}}-Q_{N_{1}}}
	{\ln\left(\frac{N_{2}}{N_{1}}\right)}
\end{equation}
and the error in the differential by
\begin{equation}
\sigma\left(\frac{dQ}{d\ln(N)}\right) \simeq
	\frac{\sqrt{\sigma^{2}(Q_{N_{2}})+\sigma^{2}(Q_{N_{1}})}}
	{\ln\left(\frac{N_{2}}{N_{1}}\right)}.
\end{equation}
where $\sigma(Q_{N})$ is the standard error in $Q_{N}$. 

To characterise the DLA clusters we measured the multipole powers,
since the corresponding multipole moments provide an orthogonal set
which may be used to totally describe the growing properties of the
clusters. In three dimensions the multipole moments are estimated
by (see \cite{jackson}, chapter 4) \begin{equation}
q_{l,m}=\frac{1}{n}\sum_{i=1,n}r_{i}^{l}Y_{l,m}(\theta_{i},\phi_{i})\end{equation}
 using $n$ probe particles which contact the cluster at $(r_{i},\theta_{i},\phi_{i})$
for $i=1$ to $n$. We normalised the multipole power as\begin{equation}
P_{l}=\frac{\sum_{m=-l}^{l}\left|q_{l,m}\right|^{2}}{(2l+1)R_{\mbox{eff}}^{2l}}\label{definition multipole power 3d}\end{equation}
 where the effective radius is in turn given by\begin{equation}
\frac{1}{R_{\mbox{eff}}}=\frac{1}{n}\sum_{i}\frac{1}{r_{i}}.\end{equation}
 The definition of $R_{\mbox{eff}}$ is such that it gives the radius of
spherical target of equivalent absorption strength to the cluster.
Note that for each measurement we used the centre of charge as origin,
meaning zero dipole moments and hence $P_{1}=0$; otherwise there
is confusion between cluster shape and drift of the cluster centre
(albeit the latter is rather negligible).

\begin{figure}
\begin{center}
\resizebox*{\columnwidth}{0.717 \columnwidth}{
\includegraphics{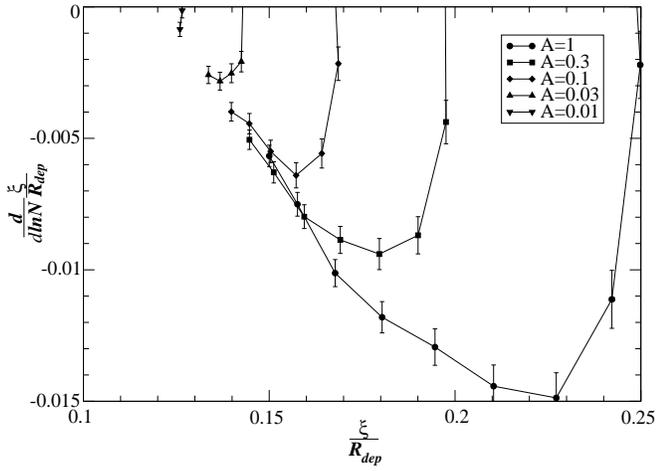}}
\end{center}
\caption[Differential plot of $\frac{\xi}{R_{dep}}$ for clusters grown in 3D.]
{Differential plot of Relative Penetration
Depth $\frac{\xi}{R_{dep}}$ against its own derivative with respect
to $\ln N$. The intercept with zero derivative indicates the asymptotic
value of $\frac{\xi}{R_{dep}}$ for infinite $N$ is given by $\left.\frac{\xi}{R_{dep}}\right|_{\infty}=0.122\pm0.002$
and the common limiting slope of the plots indicates a correction
to scaling exponent of $\nu=0.22\pm0.03$. \label{differential relative penetration 3d}
Growth at different levels of noise reduction $A\geq0.03$ is consistent
with universal values of asymptote and exponent, whilst $A=0.01$ appears
to start and remain close to the 'fixed point' value of $\frac{\xi}{R_{dep}}$.}
\end{figure}

\begin{figure*}
\parbox{0.49 \textwidth}{\raisebox{0.8cm}{
\resizebox*{0.49 \textwidth}{0.36 \textwidth}{
\includegraphics{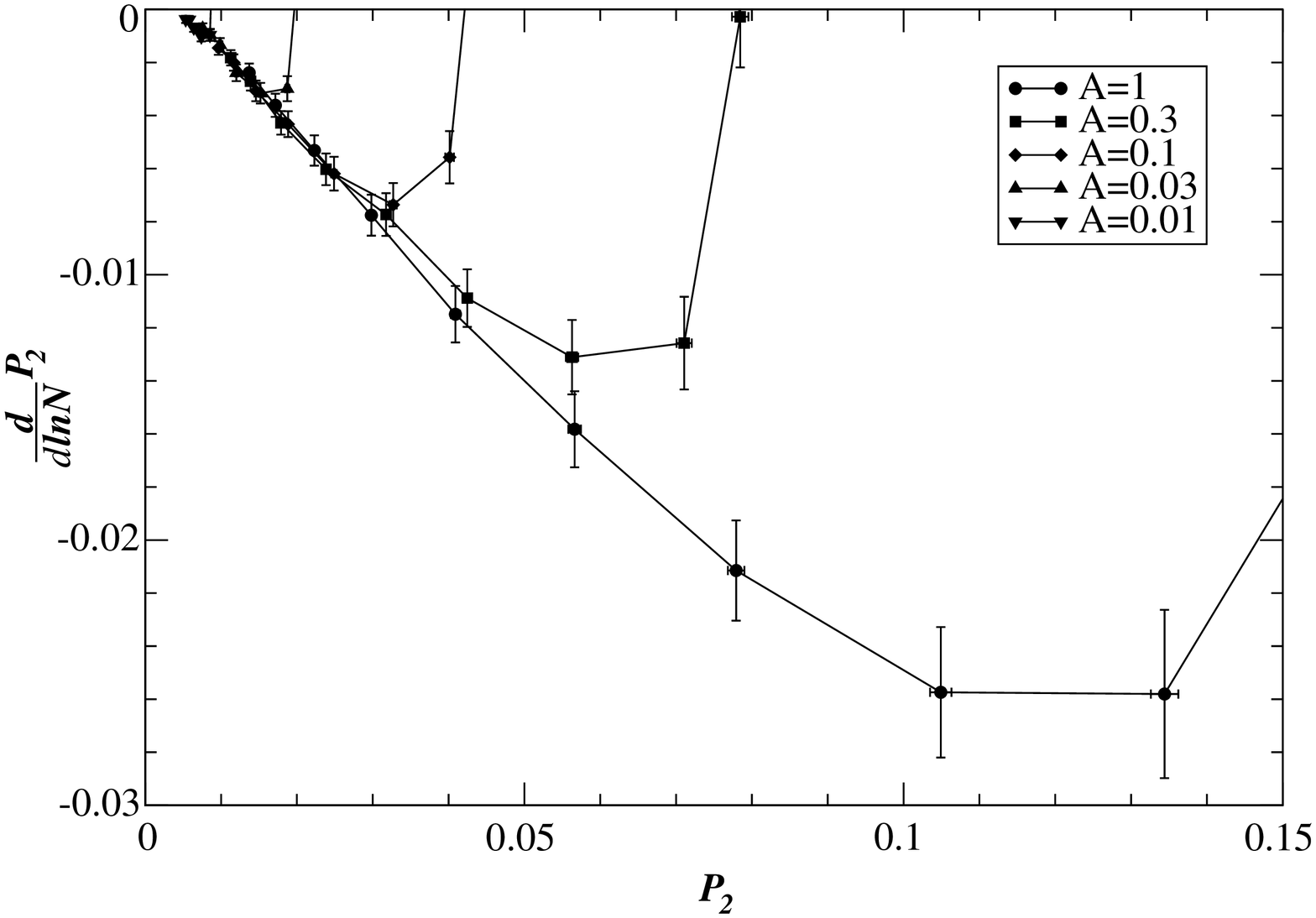}}}}
\parbox{0.49 \textwidth}{\raisebox{0.8cm}{
\resizebox*{0.49 \textwidth}{0.36 \textwidth}{
\includegraphics{figure4b.eps}}}}
\parbox{0.49 \textwidth}{\resizebox*{0.49 \textwidth}{0.36 \textwidth}{
\includegraphics{figure4c.eps}}}\hspace{0.2cm}
\parbox{0.49 \textwidth}{\resizebox*{0.49 \textwidth}{0.36 \textwidth}{
\includegraphics{figure4d.eps}}}
\caption[Differential plots for the first four (non-trivial) multipole moments $P\_2 - P\_5$ for clusters grown in 3D.]
{Differential plots for
the first four (non-trivial) multipole moments $P_{2}-P_{5}$. All
of the plots exhibit universal asymptotic values, corresponding to
the extrapolation to zero derivative, and universal limiting slope
corresponding to their correction to scaling exponent. The correction
to scaling exponents were estimated by eye as: $\nu(P_{2})=0.32\pm0.02$,
$\nu(P_{3})=0.24\pm0.03$, $\nu(P_{4})=0.26\pm0.06$, and $\nu(P_{5})=0.29\pm0.05$,
where the errors represent the maximum believable error.}
\label{differential multipoles 3d}
\end{figure*}

\begin{figure}
\begin{center}
\resizebox*{\columnwidth}{0.717 \columnwidth}{
\includegraphics{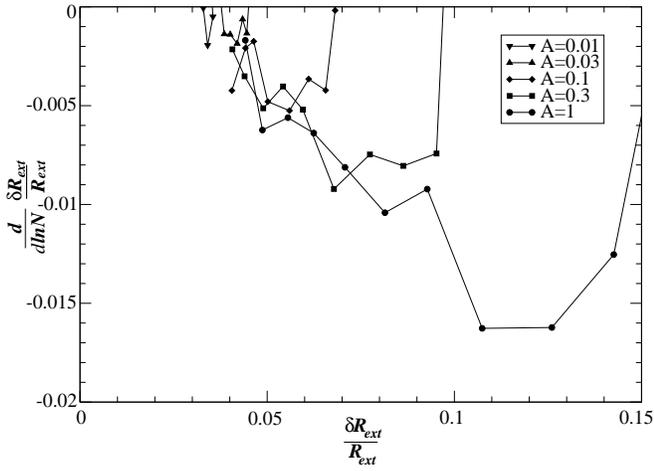}}
\end{center}
\caption[Differential plot of $\frac{\delta R_{ext}}{R_{ext}}$ for clusters grown in 3D.]
{Differential plot of the relative fluctuation
in extremal radius, $\frac{\delta R_{ext}}{R_{ext}}$. The asymptotic
value is $\left.\frac{\delta R_{ext}}{R_{ext}}\right|_{\infty}=0.032\pm0.004$
which leads to an estimate of the fixed point noise reduction of $\epsilon^{*}=0.0064\pm0.0016$.}
\label{differential relative variability 3d}
\end{figure}

The first and important feature of our differential plots is the indication
of limiting asymptotic values $Q_{\infty}$(corresponding to $\frac{dQ}{d\ln(N)}=0$)
which are universal, independent of the level of noise reduction.
This is shown for the relative penetration depth in fig.\ \ref{differential relative penetration 3d},
multipole powers, $P_{2}$ - $P_{5}$ being shown in fig.\ \ref{differential multipoles 3d},
and the relative variability of extremal cluster radius in fig.\ 
\ref{differential relative variability 3d}. The universality of the
asymptotic values of all these plots is strong indication that the
limiting distribution of cluster shape is universal.

\begin{figure}
\begin{center}
\resizebox*{\columnwidth}{0.717 \columnwidth}{
\includegraphics{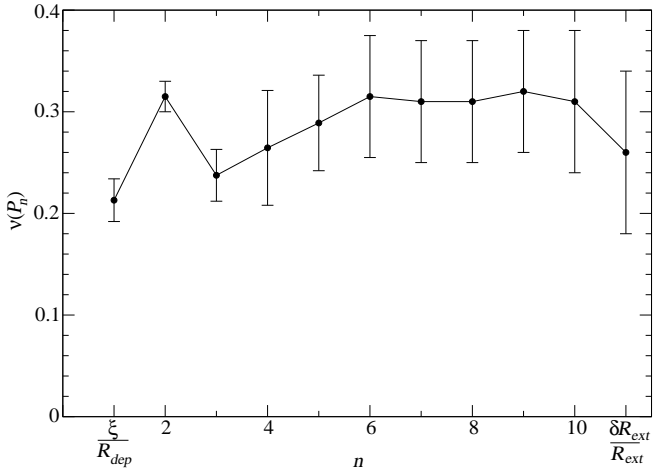}}
\end{center}
\caption[The correction to scaling exponents measured for 3D DLA clusters.]
{The correction to scaling exponents obtained from differential plots
of different multipole powers, and the two other quantities marked.
The exponent for the dipole power $P_{2}$ is significantly different
from the exponents measured for $\frac{\xi}{R_{dep}}$ and $P_{3}$.}
\label{exponents 3d}
\end{figure}

The slopes of these same plots indicate the correction to scaling
exponents, which also appear to be universal with respect to the noise
reduction. Figure \ref{exponents 3d} shows the measured values of
the correction to scaling exponents for each of the quantities plotted
(and all multipole moments measured). The exponent for the quadrupole
power $P_{2}$ is significantly different from the exponents for $\frac{\xi}{R_{dep}}$
and $P_{3}$. There is no quantity which shows a slower correction
to scaling than $\frac{\xi}{R_{dep}}$, which suggests that the result
found by Somfai \emph{et al.\ {}}\cite{somfai} in some sense also
applies to clusters grown in three dimensions. 

The values of the correction to scaling exponents are least precise
for the highest multipole moments, as these are most sensitive to
the fine structure of the cluster. Intriguingly, the asymptotic value
of the relative penetration depth for DLA clusters in three dimensions
is equal within measurement error to that of the two dimensional case:
see figure \ref{differential relative penetration 3d} here and figure
3 in \cite{ball2002}. The correction to scaling exponent for $\frac{\xi}{R_{dep}}$
is around $\frac{2}{3}$ the value of the exponent for clusters grown
in two dimensions, indicating that DLA clusters in three dimensions
are considerably slower to mature.

\section{Fixed point noise reduction}

It is clear from differential plots such as figure \ref{differential relative penetration 3d}
that the noise reduction {}``controls'' the slowest correction to
scaling. For low values of $A$ this correction to scaling is strongly
reduced, and we may write the behaviour of this correction as follows
\begin{equation}
Q_{N}=Q_{\infty}+C(A)N^{-\nu}.\end{equation}
 Other corrections to scaling need not depend on $A$, but it is quite
evident that the amplitude of the slowest correction to scaling is
strongly affected by it. If there is some value of $A$ for which
$C(A)$ is zero, then this value of the noise reduction would correspond
to the fixed point of a renormalisation scheme (see \cite{barker}).
The plots for $P_{2}-P_{4}$ suggest that the fixed point noise reduction
is $A^{f}<0.01$ and plots of $\frac{\xi}{R_{dep}},P_{5}$ are unclear
as to the value of the fixed point noise reduction. Hence, one estimates
that \begin{equation}
A^{f}\leq0.01.\label{fastest noise reduction}\end{equation}

The noise reduction of the fixed point can also be estimated from
the asymptotic properties of DLA clusters using the Barker \& Ball
\cite{barker} formula \begin{equation}
\epsilon^{*}=D^{2}\left(\frac{\delta R_{ext}}{R_{ext}}\right)^{2}\end{equation}
 where $R_{ext}$ is the extremal cluster radius (the radius of the
furthest cluster particle from the seed particle) and $\delta R_{ext}$
is the cluster to cluster variability of $R_{ext}$. From figure \ref{differential relative variability 3d},
the asymptotic value of the relative variability of extremal cluster
radius is $\left.\frac{\delta R_{ext}}{R_{ext}}\right|_{\infty}=0.032\pm0.004$.
This leads to \begin{equation}
\epsilon^{*}=0.0064\pm0.0016,\end{equation}
 which is close to the estimated value of $A^{f}$. 

\begin{figure}
\begin{center}
\resizebox*{\columnwidth}{0.717 \columnwidth}{
\includegraphics{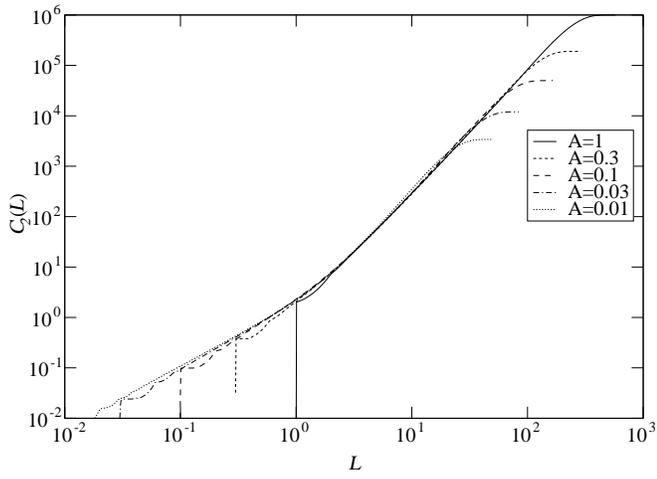}}
\end{center}
\caption[The 2 point correlation function for clusters grown in 3 dimensions.]
{The two point correlation function for clusters grown in three dimensions,
scaled by effective noise reduction factors, $\epsilon^{\mbox{eff}}$. This
noise reduction is chosen so that a data collapse is seen for small $L$.}
\label{c2 plot 3d}
\end{figure}

For a noise reduction of $A$, one would naively assume that it would
require of order $N/A$ particles to grow a cluster with the same
radius as a non-noise reduced cluster of $N$ particles. Our data
below show that this is a systematic underestimate, so that each value
of $A$ corresponds to a more severe noise reduction than expected.
Figure \ref{c2 plot 3d} shows the two point correlation function
for DLA clusters grown at different noise reductions. The graphs have
been shifted vertically so that all the curves collapse to a single
line. From the shift factors used, we estimate the effective noise
reductions to be \begin{equation}
\begin{array}{ccc}
A=1 & \epsilon^{\mbox{eff}}=1\\
A=0.3 & \epsilon^{\mbox{eff}}=0.19\\
A=0.1 & \epsilon^{\mbox{eff}}=0.05\\
A=0.03 & \epsilon^{\mbox{eff}}=0.012\\
A=0.01 & \epsilon^{\mbox{eff}}=0.0034.\end{array}\end{equation}
 Hence one concludes that the fixed point noise reduction in equation
\ref{fastest noise reduction} should be adjusted to \begin{equation}
\epsilon^{f}\leq0.0035.\end{equation}
 The values for $\epsilon_{f}$ and $\epsilon^{*}$ differ by a factor
of 2, demonstrating that the identification process is open to some
errors. If, as indicated by the results in figure \ref{exponents 3d},
a single slowest correction to scaling exponent does not control the
scaling of all parameters, then a renormalisation scheme which is
based on a single parameter (the noise reduction) may be inaccurate
and agreement between $\epsilon^{f}$ and $\epsilon^{*}$ is not expected
to be perfect.

\section{Conclusion}

The growth off-lattice of noise-reduced diffusion-limited aggregates
in three dimensions has been considered and shown to exhibit universality
with respect to noise reduction. The fractal dimension is found to
be $D=2.50\pm0.01$ which agrees with previous computational \cite{tolman}
and mean field \cite{tokuyama1984} estimates. The penetration depth
scales with the radius, and the asymptotic value of the relative penetration
depth is $\frac{\xi}{R_{dep}}=0.122\pm0.002$ which overlaps the value
found for clusters grown in two dimensions \cite{ball2002}. The convergence
of the multipole powers provides a very strong indication that DLA
cluster growth, in three dimensions and off-lattice, converges to
a universal distribution of cluster shapes.

The relative penetration depth exhibited the slowest correction to
scaling, $N^{0.22\pm0.03}$, and some multipole powers and also the
relative fluctuations in extremal radius exhibited correction to scaling
exponents which could be consistent with the same. However not all
quantities exhibit the influence of the slowest correction to scaling
and in particular the convergence to scaling of the dipole power,
$P_{2}$, is significantly faster than that of either $\frac{\xi}{R_{dep}}$
or $P_{3}$.

Reducing the input noise by factors up to 100, by growing clusters
in shallow bumps, clearly reduces the amplitude of the leading correction
to scaling. This supports in three dimensions the idea of Barker \& Ball \cite{barker}
that the intrinsic fluctuation level is the physical origin of that
slowest correction to scaling. We estimated the fixed point noise
reduction to be $\epsilon^{f}\sim0.0035$ and this is close to the
value estimated using the Barker \& Ball \cite{barker} formula in
terms of relative fluctuation in extremal radius. 

Taken together our results support the hypothesis that isotropic DLA
in three dimensions approaches a simple ensemble of statistically
self-similar clusters, with a rather slow approach to scaling which
is associated with the level of local geometric fluctuation. From
this point of view, a quantitative model of the convergence of that
fluctuation level appears to be the outstanding challenge in understanding
isotropic DLA (in any dimension). Another important challenge for
three dimensions, for which work is in progress, is the role which
material anisotropy can play \cite{goold}.

\hrulefill

The authors wish to thank Paul Meakin \& Thomas Rage for supplying
code used for growing clusters in this study, and Leonard Sander and
Ellak Somfai for their enlightening discussions. 

\bibliographystyle{apsrev}
\bibliography{../References}

\begin{thebibliography}{27}
\expandafter\ifx\csname natexlab\endcsname\relax\def\natexlab#1{#1}\fi
\expandafter\ifx\csname bibnamefont\endcsname\relax
  \def\bibnamefont#1{#1}\fi
\expandafter\ifx\csname bibfnamefont\endcsname\relax
  \def\bibfnamefont#1{#1}\fi
\expandafter\ifx\csname citenamefont\endcsname\relax
  \def\citenamefont#1{#1}\fi
\expandafter\ifx\csname url\endcsname\relax
  \def\url#1{\texttt{#1}}\fi
\expandafter\ifx\csname urlprefix\endcsname\relax\def\urlprefix{URL }\fi
\providecommand{\bibinfo}[2]{#2}
\providecommand{\eprint}[2][]{\url{#2}}

\bibitem[{\citenamefont{Witten and Sander}(1981)}]{witten}
\bibinfo{author}{\bibfnamefont{T.~A.} \bibnamefont{Witten}} \bibnamefont{and}
  \bibinfo{author}{\bibfnamefont{L.~M.} \bibnamefont{Sander}},
  \bibinfo{journal}{Physical Review Letters} \textbf{\bibinfo{volume}{47}},
  \bibinfo{pages}{1400} (\bibinfo{year}{1981}).

\bibitem[{\citenamefont{Brady and Ball}(1984)}]{brady}
\bibinfo{author}{\bibfnamefont{R.~M.} \bibnamefont{Brady}} \bibnamefont{and}
  \bibinfo{author}{\bibfnamefont{R.~C.} \bibnamefont{Ball}},
  \bibinfo{journal}{Nature} \textbf{\bibinfo{volume}{309}},
  \bibinfo{pages}{225} (\bibinfo{year}{1984}).

\bibitem[{\citenamefont{Fujikawa and Matsushita}(1989)}]{fujikawa}
\bibinfo{author}{\bibfnamefont{H.}~\bibnamefont{Fujikawa}} \bibnamefont{and}
  \bibinfo{author}{\bibfnamefont{M.}~\bibnamefont{Matsushita}},
  \bibinfo{journal}{Journal of the Physical Society of Japan}
  \textbf{\bibinfo{volume}{58 No. 11}}, \bibinfo{pages}{3875}
  (\bibinfo{year}{1989}).

\bibitem[{\citenamefont{Langer}(1980)}]{langer}
\bibinfo{author}{\bibfnamefont{J.~S.} \bibnamefont{Langer}},
  \bibinfo{journal}{Reviews of Modern Physics} \textbf{\bibinfo{volume}{52 No.
  1}}, \bibinfo{pages}{1} (\bibinfo{year}{1980}).

\bibitem[{\citenamefont{Nittmann et~al.}(1985)\citenamefont{Nittmann, Daccord,
  and Stanley}}]{nittmann}
\bibinfo{author}{\bibfnamefont{J.}~\bibnamefont{Nittmann}},
  \bibinfo{author}{\bibfnamefont{G.}~\bibnamefont{Daccord}}, \bibnamefont{and}
  \bibinfo{author}{\bibfnamefont{H.~E.} \bibnamefont{Stanley}},
  \bibinfo{journal}{Nature} \textbf{\bibinfo{volume}{314}},
  \bibinfo{pages}{141} (\bibinfo{year}{1985}).

\bibitem[{\citenamefont{Mandelbrot}(1992)}]{mandelbrot4}
\bibinfo{author}{\bibfnamefont{B.~B.} \bibnamefont{Mandelbrot}},
  \bibinfo{journal}{Physica A} \textbf{\bibinfo{volume}{191}},
  \bibinfo{pages}{95} (\bibinfo{year}{1992}).

\bibitem[{\citenamefont{Somfai et~al.}(1999)\citenamefont{Somfai, Sander, and
  Ball}}]{somfai}
\bibinfo{author}{\bibfnamefont{E.}~\bibnamefont{Somfai}},
  \bibinfo{author}{\bibfnamefont{L.~M.} \bibnamefont{Sander}},
  \bibnamefont{and} \bibinfo{author}{\bibfnamefont{R.~C.} \bibnamefont{Ball}},
  \bibinfo{journal}{Physical Review Letters} \textbf{\bibinfo{volume}{83 No.
  26}}, \bibinfo{pages}{5523} (\bibinfo{year}{1999}).

\bibitem[{\citenamefont{Tang}(1985)}]{tang}
\bibinfo{author}{\bibfnamefont{C.}~\bibnamefont{Tang}},
  \bibinfo{journal}{Physical Review A} \textbf{\bibinfo{volume}{31 No. 3}},
  \bibinfo{pages}{1977} (\bibinfo{year}{1985}).

\bibitem[{\citenamefont{Ball}(1985)}]{ball2}
\bibinfo{author}{\bibfnamefont{R.~C.} \bibnamefont{Ball}},
  \bibinfo{journal}{Physica A} \textbf{\bibinfo{volume}{140}},
  \bibinfo{pages}{62} (\bibinfo{year}{1985}).

\bibitem[{\citenamefont{Barker and Ball}(1990)}]{barker}
\bibinfo{author}{\bibfnamefont{P.~W.} \bibnamefont{Barker}} \bibnamefont{and}
  \bibinfo{author}{\bibfnamefont{R.~C.} \bibnamefont{Ball}},
  \bibinfo{journal}{Physical Review A} \textbf{\bibinfo{volume}{42}},
  \bibinfo{pages}{6289} (\bibinfo{year}{1990}).

\bibitem[{\citenamefont{Hastings and Levitov}(1998)}]{hastings}
\bibinfo{author}{\bibfnamefont{M.~B.} \bibnamefont{Hastings}} \bibnamefont{and}
  \bibinfo{author}{\bibfnamefont{L.~S.} \bibnamefont{Levitov}},
  \bibinfo{journal}{Physica D} \textbf{\bibinfo{volume}{116}},
  \bibinfo{pages}{244} (\bibinfo{year}{1998}).

\bibitem[{\citenamefont{Ball et~al.}(2002)\citenamefont{Ball, Bowler, Sander,
  and Somfai}}]{ball2002}
\bibinfo{author}{\bibfnamefont{R.~C.} \bibnamefont{Ball}},
  \bibinfo{author}{\bibfnamefont{N.~E.} \bibnamefont{Bowler}},
  \bibinfo{author}{\bibfnamefont{L.~M.} \bibnamefont{Sander}},
  \bibnamefont{and} \bibinfo{author}{\bibfnamefont{E.}~\bibnamefont{Somfai}},
  \bibinfo{journal}{Physical Review E} \textbf{\bibinfo{volume}{66}},
  \bibinfo{pages}{026109} (\bibinfo{year}{2002}).

\bibitem[{\citenamefont{Tolman and Meakin}(1989)}]{tolman}
\bibinfo{author}{\bibfnamefont{S.}~\bibnamefont{Tolman}} \bibnamefont{and}
  \bibinfo{author}{\bibfnamefont{P.}~\bibnamefont{Meakin}},
  \bibinfo{journal}{Physical Review A} \textbf{\bibinfo{volume}{40 No. 1}},
  \bibinfo{pages}{428} (\bibinfo{year}{1989}).

\bibitem[{\citenamefont{Meakin}(1983{\natexlab{a}})}]{meakin2}
\bibinfo{author}{\bibfnamefont{P.}~\bibnamefont{Meakin}},
  \bibinfo{journal}{Physical Review A} \textbf{\bibinfo{volume}{27 No. 3}},
  \bibinfo{pages}{1495} (\bibinfo{year}{1983}{\natexlab{a}}).

\bibitem[{\citenamefont{Meakin}(1983{\natexlab{b}})}]{meakin5}
\bibinfo{author}{\bibfnamefont{P.}~\bibnamefont{Meakin}},
  \bibinfo{journal}{Physical Review A} \textbf{\bibinfo{volume}{27 No. 1}},
  \bibinfo{pages}{604} (\bibinfo{year}{1983}{\natexlab{b}}).

\bibitem[{\citenamefont{Sander et~al.}(1983)\citenamefont{Sander, Cheng, and
  Richter}}]{sander3}
\bibinfo{author}{\bibfnamefont{L.~M.} \bibnamefont{Sander}},
  \bibinfo{author}{\bibfnamefont{Z.~M.} \bibnamefont{Cheng}}, \bibnamefont{and}
  \bibinfo{author}{\bibfnamefont{R.}~\bibnamefont{Richter}},
  \bibinfo{journal}{Physical Review B} \textbf{\bibinfo{volume}{28 No. 11}},
  \bibinfo{pages}{6394} (\bibinfo{year}{1983}).

\bibitem[{\citenamefont{Gagne and Kroger}(1996)}]{gagne}
\bibinfo{author}{\bibfnamefont{R.}~\bibnamefont{Gagne}} \bibnamefont{and}
  \bibinfo{author}{\bibfnamefont{H.}~\bibnamefont{Kroger}},
  \bibinfo{journal}{Chaos, Solitons and Fractals} \textbf{\bibinfo{volume}{7
  No. 1}}, \bibinfo{pages}{125} (\bibinfo{year}{1996}).

\bibitem[{\citenamefont{Schwarzer
  et~al.}(1992{\natexlab{a}})\citenamefont{Schwarzer, Havlin, and
  Stanley}}]{schwarzer2}
\bibinfo{author}{\bibfnamefont{S.}~\bibnamefont{Schwarzer}},
  \bibinfo{author}{\bibfnamefont{S.}~\bibnamefont{Havlin}}, \bibnamefont{and}
  \bibinfo{author}{\bibfnamefont{H.~E.} \bibnamefont{Stanley}},
  \bibinfo{journal}{Physica A} \textbf{\bibinfo{volume}{191}},
  \bibinfo{pages}{117} (\bibinfo{year}{1992}{\natexlab{a}}).

\bibitem[{\citenamefont{Schwarzer
  et~al.}(1992{\natexlab{b}})\citenamefont{Schwarzer, Wolf, Havlin, Meakin, and
  Stanley}}]{schwarzer1}
\bibinfo{author}{\bibfnamefont{S.}~\bibnamefont{Schwarzer}},
  \bibinfo{author}{\bibfnamefont{M.}~\bibnamefont{Wolf}},
  \bibinfo{author}{\bibfnamefont{S.}~\bibnamefont{Havlin}},
  \bibinfo{author}{\bibfnamefont{P.}~\bibnamefont{Meakin}}, \bibnamefont{and}
  \bibinfo{author}{\bibfnamefont{H.~E.} \bibnamefont{Stanley}},
  \bibinfo{journal}{Physcial Review A} \textbf{\bibinfo{volume}{46 No. 6}},
  \bibinfo{pages}{R3016} (\bibinfo{year}{1992}{\natexlab{b}}).

\bibitem[{\citenamefont{Davidovitch et~al.}(2001)\citenamefont{Davidovitch,
  Jensen, Levermann, Mathiesen, and Procaccia}}]{jensen}
\bibinfo{author}{\bibfnamefont{B.}~\bibnamefont{Davidovitch}},
  \bibinfo{author}{\bibfnamefont{M.~H.} \bibnamefont{Jensen}},
  \bibinfo{author}{\bibfnamefont{A.}~\bibnamefont{Levermann}},
  \bibinfo{author}{\bibfnamefont{J.}~\bibnamefont{Mathiesen}},
  \bibnamefont{and}
  \bibinfo{author}{\bibfnamefont{I.}~\bibnamefont{Procaccia}},
  \bibinfo{journal}{Physical Review Letters} \textbf{\bibinfo{volume}{86 No.
  16}}, \bibinfo{pages}{164101} (\bibinfo{year}{2001}).

\bibitem[{\citenamefont{Batchelor and Henry}(1996)}]{batchelor}
\bibinfo{author}{\bibfnamefont{M.~T.} \bibnamefont{Batchelor}}
  \bibnamefont{and} \bibinfo{author}{\bibfnamefont{B.~I.} \bibnamefont{Henry}},
  \bibinfo{journal}{Physica A} \textbf{\bibinfo{volume}{233}},
  \bibinfo{pages}{905} (\bibinfo{year}{1996}).

\bibitem[{\citenamefont{Vespignani and Pietronero}(1991)}]{vespignani4}
\bibinfo{author}{\bibfnamefont{A.}~\bibnamefont{Vespignani}} \bibnamefont{and}
  \bibinfo{author}{\bibfnamefont{L.}~\bibnamefont{Pietronero}},
  \bibinfo{journal}{Physica A} \textbf{\bibinfo{volume}{173 No. 1-2}},
  \bibinfo{pages}{1} (\bibinfo{year}{1991}).

\bibitem[{\citenamefont{Ball and Brady}(1985)}]{ball}
\bibinfo{author}{\bibfnamefont{R.~C.} \bibnamefont{Ball}} \bibnamefont{and}
  \bibinfo{author}{\bibfnamefont{R.~M.} \bibnamefont{Brady}},
  \bibinfo{journal}{Journal of Physics A} \textbf{\bibinfo{volume}{18}},
  \bibinfo{pages}{L809} (\bibinfo{year}{1985}).

\bibitem[{\citenamefont{Bowler}(2001)}]{bowler2001}
\bibinfo{author}{\bibfnamefont{N.~E.} \bibnamefont{Bowler}}, Ph.D. thesis,
  \bibinfo{school}{Warwick University} (\bibinfo{year}{2001}).

\bibitem[{\citenamefont{Jackson}(1975)}]{jackson}
\bibinfo{author}{\bibfnamefont{J.~D.} \bibnamefont{Jackson}},
  \emph{\bibinfo{title}{Classical Electrodynamics (second edition)}}
  (\bibinfo{publisher}{John Wiley \& Sons}, \bibinfo{address}{Chichester},
  \bibinfo{year}{1975}).

\bibitem[{\citenamefont{M.~Tokuyama}(1984)}]{tokuyama1984}
\bibinfo{author}{\bibfnamefont{K.~K.} \bibnamefont{M.~Tokuyama}},
  \bibinfo{journal}{Physics Letters A} \textbf{\bibinfo{volume}{100 No. 7}},
  \bibinfo{pages}{337} (\bibinfo{year}{1984}).

\bibitem[{\citenamefont{Goold et~al.}()\citenamefont{Goold, Ball, and
  Somfai}}]{goold}
\bibinfo{author}{\bibfnamefont{N.~R.} \bibnamefont{Goold}},
  \bibinfo{author}{\bibfnamefont{R.~C.} \bibnamefont{Ball}}, \bibnamefont{and}
  \bibinfo{author}{\bibfnamefont{E.}~\bibnamefont{Somfai}},
  \emph{\bibinfo{title}{New simulations of diffusion controlled growth}},
  \bibinfo{note}{poster presented at Institute of Physics, Condensed Matter and
  Materials Conference, University of Warwick, April 2004}.

\end{thebibliography}

\newpage

\end{document}